\documentclass[12pt,fleqn]{article}
\usepackage{verbatim,cite,amssymb,amsmath,graphics,amsthm,subfigure}
\numberwithin{equation}{section}

\theoremstyle{definition}

\theoremstyle{definition}

\renewcommand{\H}{{\cal H}}

\newcommand{\<}{\langle}
\newcommand{\be}{\begin{equation}}
\renewcommand{\>}{\rangle}

\begin{document}

\title{Time Asymmetry in Quantum Physics - I. Theoretical Conclusion
  from Resonance and Decay-Phenomenology} \author{
  A.~Bohm$^{\star}$, H.~Kaldass$^{\dagger}$, S.~Komy$^{\ddagger}$\\
  $^{\star}$\textit{\footnotesize CCQS, Physics Department,
    University of Texas, Austin, Texas 78712}\\
  $^{\dagger}$\textit{\footnotesize Arab Academy of Science and
    Technology, El-Horia, Heliopolis,
    Cairo, Egypt}\\
  $^{\ddagger}$\textit{\footnotesize Mathematics Department, Helwan
    University, Cairo, Egypt}\\
  {\footnotesize Emails: bohm@physics.utexas.edu,
    hani@ifh.de, sol\_komy@yahoo.com }} \date{} \maketitle
\begin{abstract}
  It is explained how the unification of resonance and decay phenomena
  into a consistent mathematical theory leads to quantum mechanical
  time-asymmetry. This provides the theoretical basis for a subsequent
  paper~II in which the interpretation and experimental demonstration
  of this time-asymmetry is discussed.
\end{abstract}
\section{Introduction}
Within the framework of traditional quantum theory one does not have a
consistent theory of resonance and decay phenomena. One has various
empirical concepts and useful methods, but, many puzzles, questions
and contradictions remain. In non-relativistic physics, one has at
least a generally accepted calculational scheme, the Weisskopf-Wigner
approximation \cite{1}.

In the relativistic domain, one cannot even agree upon an {\em
  approximate} description of resonances. One is not sure whether it
makes sense at all to describe resonances as separate entities which
can be characterized by two well defined quantities, the mass and the
width. In particular, recently in connection with the $Z$- and
$W$-bosons it was mentioned that the definition of $M$ and $\Gamma$ is
just a convention and as long as this was done consistently, more or
less any parametrization of the complex pole position was acceptable.
With this argument one then justifies the use of the old
parametrization of the complex pole position in terms of the non-gauge
invariant on-the-mass-shell mass $M_Z$ and width $\Gamma_Z$~\cite{1a}.

In non-relativistic quantum mechanics, on the basis of the
Weisskopf-Wigner approximation \cite{1} the width $\Gamma$ of a
Breit-Wigner energy distribution \eqref{2} is connected to the inverse
lifetime $\tau$.  The Weisskopf-Wigner approximate methods provide
only a vague and approximate relation \cite{2}, see \eqref{6.2} below.
If $\Gamma = \frac{\hbar}{\tau}$ could be established as an exact
relation, then one could use this relation as the criterion for the
right definition of the width $\Gamma$. With the width precisely
defined, this would then also define the mass $M$ and therewith
uniquely fix the two parameters $(M,\Gamma)$ of a relativistic
resonance.  The first task, therefore, is to obtain a consistent
quantum theory that unifies resonance scattering and decay phenomena,
such that a relation $\Gamma = f(\tau)$, preferably, $\Gamma =
\frac{\hbar}{\tau}$ is obtained as a result of this theory.

Such a theory cannot be obtained within the framework of conventional
quantum mechanics using the Hilbert space axiom, because -- as is well
known -- the Hilbert space does not contain a vector (or a state) with
exponential time evolution. To obtain exponential Born probabilities
which are needed to define the lifetime $\tau$ of exponential decay,
one has to use generalized vectors, like e.g., Dirac kets, but still
more generalized.  Whereas Dirac kets $|E\>$ are eigenkets (i.e.,
continuous functionals) with eigenvalues from the continuous energy
spectrum, these new generalized eigenvectors are eigenkets with
complex eigenvalue $|E_R-i\Gamma/2^-\>$; for this reason, we call them
Gamow kets. In mathematical terms, they are continuous functionals on
a Hardy space $\Phi_-$, whereas a Dirac ket, if defined at all, is
mathematically defined as a continuous functional on a Schwartz space,
\textit{cf.}~\cite{4}.

To obtain these Gamow kets with Breit-Wigner width $\Gamma$ and
lifetime $\tau = \frac{\hbar}{\Gamma}$, we have to modify just one of
the traditional axioms of quantum mechanics, the Hilbert space axiom,
and replace it with the Hardy space axiom. The Hardy space axiom
mathematically distinguishes between prepared in-states and detected
out-observables (usually miss-named out-states). Describing the set of
in-states by $\Phi_-\subset\H$ and the set of detected out-observables
by $\Phi_+\subset\H$, both dense in the same Hilbert space $\H$ for a
particular quantum system, one obtains a consistent and exact theory
that unifies quantum resonances and decay: From the $S$-matrix pole
definition of a resonance, one obtains a Gamow ket (functional on
$\Phi_-$) with Breit-Wigner (Lorentzian) energy distribution and
exponential time evolution. This Gamow ket represents the resonance
per se (without background).

But, one also predicts as a mathematical consequence of the new Hardy
space boundary conditions, $\phi^+(t)\in\Phi_-$ for the Schr\"odinger
equation and $\psi^-(t)\in\Phi_+$ for the Heisenberg equation, a
time-asymmetric semigroup evolution. The semigroup is in contrast to
the reversible unitary group evolution of Hilbert space quantum
mechanics. The semigroup evolution introduces a new concept into
quantum mechanics, the semigroup time, $t_0 = 0$, which does not exist
in conventional quantum mechanics, where the time extends over
$-\infty < t < \infty$. The quantum mechanical ``beginning of time''
$t_0 (> -\infty)$ and the experimental demonstration of it will be
discussed in a subsequent paper. As a preparation of this, we explain
in the present paper why quantum mechanical time asymmetry $t_0 = 0 <
t < \infty$ is a consequence of the exact equality $\tau =
\frac{\hbar}{\Gamma}$.

\section{Resonances and Decaying States}
Resonances and decaying states are widely believed to be different
manifestations of the same entities.

Resonances ${\cal R}$ appear as intermediate states of scattering
processes
\begin{equation}
 1+2\to {\cal R} \to 3+4, \quad \text{for example}\quad e^+e^-\to Z^0
 \to \mu^+\mu^-\,,
\label{1}
\end{equation}
when the scattering cross section of angular momentum $j$,
$|a^{BW}_j(E)|^2$ is described by a Breit-Wigner energy distribution,
(also called Lorentzian):
\begin{equation}
a^{BW}_j=\frac{r_\eta}{E-(E_R-i\frac{\Gamma}{2})};\quad 0\le
E<\infty\,.\label{2}
\end{equation}

Resonances are thus characterized by the angular momentum $j$ and by
the resonance energy $E_R$ (or resonance mass $M$ in the relativistic
case) and the resonance width $\Gamma$.

Usually the Lorentzian \eqref{2} is not sufficient to fit the
experimental cross section (and asymmetry data) for processes like
\eqref{1}; in addition to the resonance amplitude \eqref{2} there is
always a background amplitude $B_j$ and the partial cross section of
angular momentum $j$ is fitted to:
\begin{equation}
\label{2.2}
\sigma_j(E)\sim |a_j(E)|^2 = |a_j^{BW}(E) + B_j(E)|^2\,.
\end{equation}

Decaying states $\phi^D(t)$ are considered as the starting points of
an exponential time evolution. Decaying states are observed in
processes like
\begin{equation}
D\to \eta_1,\eta_2,\cdots,\quad \textrm{e.g.,}\,\,\, K^0_S\to
\pi^+\pi^-\,,\,\pi^0\pi^0
\label{3}
\end{equation}
where $\eta_1,\eta_2,\cdots$ denote different decay products (or decay
channels). The decay products, or more precisely the properties of the
decay products $\eta$ are described by the out ``state'' vector
$\psi_\eta$, or out-observable $\Lambda_\eta =
|\psi_\eta\>\<\psi_\eta|$.

The decaying state $\phi^D$ is characterized by $(E_D, 1/\tau\equiv
R)$ (or $(M, 1/\tau\equiv R)$) where $\tau$ is the lifetime (for the
relativistic case, in the rest frame) and $R$ is the total initial
decay rate. The lifetime $\tau$ is measured by fitting the
experimental counting rate, $\frac{1}{N}\frac{\Delta N_\eta
  (t)}{\Delta t}$, for any decay product $\eta$ to the partial decay
rate $R_\eta(t)$ (the intensity of the $\eta$ emission as a function
of time), for which one assumes the empirical exponential law
\begin{equation}
\frac{1}{N}\frac{\Delta N_\eta(t_i)}{\Delta t_i}\approx R_\eta(t) = 
R_\eta(0)e^{-t/\tau}=R_\eta(0)e^{-Rt}\,.\label{4}
\end{equation}
Here $R(t)=\sum_\eta R_\eta(t)$ is the total decay rate and $R = R(0)$
is the total initial decay rate; $\Delta N_\eta(t_i)$ is the number of
the decay products $\eta$ registered by the $\eta$-detector during the
time interval $\Delta t_i$ around $t_i$.

The theoretical decay rates $R_\eta(t)$ and the probabilities
$P_\eta(t)$, are according to a fundamental axiom of quantum
mechanics, given by the quantum mechanical Born probabilities of the
observable $\Lambda_\eta$ in the (decaying) state $\phi^D(t)$:
\begin{equation}
P_\eta(t)=Tr(\Lambda_\eta|\phi^D(t)\>\<\phi^D(t)|)=|\<\psi_\eta|\phi^D(t)\>|^2
\label{5}
\end{equation}
where $\Lambda_\eta = |\psi_\eta\>\<\psi_\eta|$ is the projection
operator on the subspace of properties of the decay products which are
registered by the $\eta$-detector.  The partial decay rates (also
misleadingly called partial widths when multiplied by $\hbar$,
$\Gamma_\eta=\hbar R_\eta(0)$) are the time derivatives of the
probabilities $P_\eta(t)$.
\begin{equation}
R_\eta(t)=\frac{d}{dt}P_\eta(t).\label{6}
\end{equation}
The experimental definition of the lifetime $\tau$, given by \eqref{4}
and the relation between the total initial decay rate and the lifetime
$\tau$, $R=\frac{1}{\tau}$, is based on the validity of the
exponential law (\ref{4}) for the rate $R(t)$, and therefore also on
the exponential law for the decay probabilities $P_\eta(t)$. {\em If
  the exponential law holds}, the total decay rate $R$ and the inverse
lifetime are the same\footnote{The average lifetime is defined as the
  average value of the time intervals $\Delta t_i$ that any one of the
  particles $D$ survives:
\begin{equation}
\label{star}
\frac{1}{N_D}\sum_i N_D(\Delta t_i)\Delta t_i \approx
\frac{1}{N_D}\int N_D(t)dt = \tau
\end{equation}
where $N_D = N_D(0) = N_D(t) + \sum_\eta N_\eta(t)$ is the number of
decaying particles at $t = 0$, $N_D(t)$ is the number of decaying
particles at time $t$, and $N_\eta(t)$, $\eta =
\eta_1\,,\eta_2\,,\cdots$, are the numbers of decay products $\eta$ at
time $t$. If the exponential law ${\cal P}_D(t) \approx
\frac{N_D(t)}{N_D} = e^{-Rt}$ holds, and only if the exponential law
holds, is the average lifetime given by $\frac{1}{R} = \tau$.}.
However, the inverse lifetime $1/\tau$ of (\ref{4}) and the width
$\Gamma$ of the Breit-Wigner energy distribution (\ref{2}) are
conceptually and experimentally different quantities. The width
$\Gamma$ is determined experimentally from the fit of the scattering
data to the Lorentzian lineshape (\ref{2}) of a resonance scattering
experiment \eqref{1}.  The lifetime $\tau=1/R$ is determined
experimentally from a fit to the exponential time dependence of the
counting rate of the decay products (\ref{4}).

Nevertheless, one often calls the calculated quantity
$\Gamma^{\textrm{calc}}=\hbar R=\hbar /\tau$ with $\tau$ measured by
(\ref{4}), also the width of the decaying particle, and the $\hbar
R_\eta\equiv \Gamma_\eta$ are usually called the partial widths.

Within traditional quantum mechanics (using the Hilbert space axiom)
one cannot derive the equality of $\Gamma$ in \eqref{2} and $\hbar R =
\hbar/\tau$ in \eqref{4} or any other relationship between the width
$\Gamma$ and the inverse lifetime $\frac{\hbar}{\tau}$. One can also
not derive the exponential law (\ref{4}) from (\ref{5}) using
\eqref{6}, because there is no vector $\phi^D(t)$ in Hilbert space for
which the right hand side of \eqref{6} with \eqref{5} would have an
exponential time dependence.

The idea that $\tau$ and $\hbar/\Gamma$ are the same or are at least
approximately equal is based on the Weisskopf-Wigner approximation
methods \cite{1,2}. In the monograph \cite{2} one considers a prepared
state $\phi^D$, which has a Breit-Wigner energy wave function
(\ref{2}).  Then one calculates the probability ${\cal P}_D(t)$ for
finding this state $\phi^D$ at a time $t$ and obtains (section $8.2$
of \cite{2}):
\begin{equation}
\label{6.2}
{\cal P}_D(t) = e^{-\Gamma t/\hbar} + \Gamma\times\text{small terms}\,.
\end{equation}
Neglecting the small terms one can conclude for the average lifetime
of \eqref{star} that $\tau \approx \frac{\hbar}{\Gamma}$.

Using these kind of approximate methods a number of important
empirical notions have been introduced over the years: decaying Gamow
vectors \cite{8} with complex energy eigenvalues, Breit-Wigner
(Lorentzian) resonance amplitudes \cite{10a}, the Lippmann-Schwinger
in- and out-plane wave states \cite{10b}, the analytically continued
$S$-matrix and its resonance poles \cite{10c}. While these methods
provided a means to perform calculations leading to results which
agreed with the experiments to a satisfactory degree of accuracy, they
also led to many puzzles and contradictions: complex energy versus the
self-adjointness of the Hamiltonian; exponential decay law versus the
deviation from the exponential time evolution for any vector in the
Hilbert space; \cite{9c} exponential catastrophe versus the unitary
(reversible) time evolution \cite{10e}; and causality versus the
semi-boundedness of the Hamiltonian \cite{9b}.

The conclusion is that an exact, mathematically consistent theory of
resonance scattering and decay does not exist \cite{3}.  Neither
Breit-Wigner resonances nor exponentially decaying Gamow states are
possible within the frame of the traditional Hilbert space axiom of
quantum mechanics. Furthermore, the decay of excited atoms \cite{6a}
and of elementary particles \cite{6b} is a time asymmetric (sometimes
also called irreversible) process.  But time evolution in Hilbert
space is always time symmetric since the Schroedinger and Heisenberg
equations lead to the unitary (``reversible'') time evolution
(Stone-von Neumann theorem when solved under the Hilbert space
boundary condition) \cite{8a}.

\section{Modification of one traditional axiom of quantum theory}
Therefore a modification of the Hilbert space theory of quantum
physics is needed. This modification started with Dirac's kets for
which Schwartz created his theory of distributions, and which
Gel'fand, Maurin and their schools generalized to the Rigged Hilbert
Space theory, to prove the general Dirac basis vector expansion as the
Nuclear Spectral Theorem \cite{8b,8c}.  The quantum theory of
resonances and decay and of asymmetric time evolution requires
particular versions of Rigged Hilbert Spaces, the pair in which the
base spaces are Hardy spaces \cite{7,5,9}.

Dirac kets $|E\>$ and Dirac's $\delta$-distribution are well accepted
entities that lie beyond the mathematics of the Hilbert space. Only
few physics books give their mathematical definition \cite{4}: the
$|E\>$ are defined as functionals over the abstract Schwartz space and
$\delta(E-E_0)$ are functionals on the space of smooth rapidly
decreasing functions (Schwartz space functions) $S|_{{\mathbb R}_+}$.
The energy wave functions are Schwartz space functions:
$\phi(E)=\<E|\phi\>\in S|_{\mathbb{R}_+}$.

This means Dirac kets and Dirac's formulation of quantum mechanics
requires a Gel'fand triplet of spaces \cite{5}
\begin{equation}
\Phi\subset \mathcal{H} \subset \Phi^\times\quad \textrm{with}\quad
\phi\in \Phi,
\quad|E\>\in \Phi^\times\,,\label{8}
\end{equation}
where one chooses for $\Phi$ the abstract Schwartz space.

The same is expected of the other generalized vectors of scattering
and decay theory, like the Lippmann-Schwinger kets $|E^-\>$ and
$|E^+\>$. But these cannot be functionals over the Schwartz space
because of the infinitesimal $\mp i\epsilon$ in the Lippmann-Schwinger
equation, which is a means of formulating two distinct (outgoing and
incoming) boundary conditions. To define kets which allow analytic
continuation into the complex energy plane, and therewith the
formulation of outgoing and incoming boundary conditions, requires
that the energy wave functions: $\phi^+(E)=\<^+E|\phi^+\>$ and
$\psi^-(E)=\<^-E|\psi^-\>$ be ``better'' than Schwartz space
functions.  They must be functions that can be analytically continued
into the upper (for $\<^-E|\psi^-\>$) and lower (for $\<^+E|\phi^+\>$)
complex energy plane (of the second Riemann sheet of the analytically
continued $S$-matrix where the resonance poles are located).

Therefore we make the new Hardy space hypothesis:

The prepared in-states $\phi^+$ (experimentally given by the
preparation apparatus) are described by
\begin{equation}
\{\phi^+\}=\Phi_- \subset \mathcal{H}\subset \Phi^\times_- \,,\qquad
|E^+\>\in\Phi_-^\times\,,\label{9+}
\end{equation}
and the detected out-states, or precisely out-observables $\psi^-$
(because they are experimentally given by the detector) are described
by
\begin{equation}
\{\psi^-\}=\Phi_+\subset \mathcal{H}\subset \Phi^\times_+ \,,\qquad
|E^-\>\in\Phi_+^\times\,.\label{9-}
\end{equation}
The two spaces $\Phi_+$ and $\Phi_-$ are two different Hardy
sub-spaces (analytic in the upper and lower energy semi-plane
respectively) which are dense in the same Hilbert space $\mathcal{H}$.
This means the energy wavefunctions $\phi^+(E) = \<^+E|\phi^+\>$ of
the in-state $\phi^+$ and the energy wavefunctions $\psi^-(E) =
\<^-E|\psi^-\>$ of the out-observable $\psi^-$ are those Schwartz
space functions of $E$ which can be analytically continued into the
lower complex semi-plane for $\phi^+(E)$ and into the upper complex
semi-plane for $\psi^-(E)$, and which vanish rapidly enough at the
infinite semicircle; they are in the intersections of the Schwartz
space ${\cal S}$ and the Hardy spaces $\H_\mp^2$ \cite{5,9}:
\begin{align}
\tag{\ref{9+}$a$}\label{8a}
\<^+E|\phi^+\>&\equiv \phi^+(E)\in{\cal S}\cap{\cal H}_-^2|_{\mathbb{R}_+}\,,\\
\tag{\ref{9-}$a$}\label{9a}
\<^-E|\psi^-\>&\equiv \psi^-(E)\in{\cal S}\cap{\cal
  H}_+^2|_{\mathbb{R}_+}\Rightarrow
\overline{\<^-E|\psi^-\>} \equiv \<\psi^-|E^-\>\in{\cal S}\cap{\cal
  H}_-^2|_{\mathbb{R}_+}\,.
\end{align}
The values of the Hardy function $\phi^+(z)$ and
$\overline{\psi^-(E)}$ in the lower complex semi-plane second sheet
${\mathbb C}_-$ including the negative real axis are completely
determined from the values at the positive real axis ${\mathbb R}_+$
(the spectrum of the Hamiltonian).

The new hypothesis \eqref{9+} \eqref{9-} accounts for the fact that
the states $\phi^+$ and the observable $|\psi^-\>\<\psi^-|$ , which
are experimentally defined by different parts of the experiment
($\phi^+\in \Phi_-$ by the preparation apparatus for the in-state and
$\psi^-\in \Phi_+$ by the detector), are also represented
mathematically by different "parts" of the Hilbert space
$\mathcal{H}$. Different "parts" here means different {\em dense}
subspaces of $\mathcal{H}$\footnote{like the rational numbers being a
  dense subset of the real numbers}. This means, as long as one
considers only algebraic notion, leaving notions like convergence or
completeness aside, then the Hilbert space axiom
$\{\phi^+\}=\{\psi^-\}=\mathcal{H}$ and the Schwartz space axiom
$\{\phi^+\}=\{\psi^-\}=\Phi\subset\mathcal{H}$ and the new hypothesis
(\ref{9+}, \ref{9-}) "are all same".

By direct observation it is also difficult to distinguish between the
hypothesis (\ref{8}) and the new hypothesis (\ref{9+}, \ref{9-}). The
assumption (\ref{8}) would mean that the detector efficiency
$|\psi^-(E)|^2$ and the energy distribution of the beam
$|\phi^+(E)|^2$ are described by smooth functions, and the new
hypothesis (\ref{9+}, \ref{9-}) means that the detector is described
only by smooth functions $\{\psi^-(E)\}$ that can be analytically
continued into the upper complex energy semi-plane $\mathbb{C}_+$, and
the preparation apparatus is described only by those smooth functions
$\phi^+(E)$ which can be analytically continued into the lower complex
semi-plane $\mathbb{C}_-$.  Mathematically this analyticity
requirement makes a significant difference, and if one takes the
Fourier transform, one obtains from the analyticity hypothesis of
$\phi^+(E)$ and $\psi^-(E)$ the time asymmetry $t\ge 0$ and causality.

Also, since $\Phi_\mp\subset \Phi$ and therefore
$\Phi^\times_\mp\supset \Phi^\times$, there are more generalized
eigenvectors (kets) under the hypothesis (\ref{9+}, \ref{9-}) than the
Dirac kets $|E\>$ of (\ref{8}).  In particular there are the analytic
continuation of the bras $\<^-E|$, $E\in\mathbb{R}_+$, into the upper
complex plane $\mathbb{C}_+$, and therefore of the $|E^-\>$ into the
lower complex plane, second sheet $\mathbb{C}_-$ and the
\begin{equation}
|z^-\>\in \Phi^\times_+\qquad \textrm{are defined for all}\,\,
 z\in\mathbb{C}_-\,,\label{10}
\end{equation}
as long as these $z\in\mathbb{C}_-$ are not singularities of the
analytically continued $S$-matrix $S(z)$.  Similarly the
$\<^+E|\phi^+\>$, $E\in\mathbb{R}_+$, can be analytically continued
into the lower complex plane, second sheet, and bras $\<^+E|$ are thus
defined also for complex $z\in{\mathbb C}_-$
\begin{equation}
\<^+z|\in \Phi^\times_-\qquad \textrm{for}\,\, z\in\mathbb{C}_-\label{11}
\end{equation}
as long as these $z\in\mathbb{C}_-$ are not singularities of the
analytically continued $S(z)$ .  Here $\mathbb{C}_-$ refers of the
lower complex semi-plane of the second sheet of the Riemann surface of
the $S$-matrix.

In addition to the continuum of kets $|z^-\>\in\Phi_+^\times$ obtained
by analytic continuation from the $|E^-\>$ there are also the kets
$|z^-_R\>\in \Phi^\times_+$ which correspond to the discrete sets of
first order poles at $z_R = E_R-i\Gamma/2$ of the $S$-matrix. These
kets we shall call Gamow vectors; they are the central concepts of the
theory of resonance scattering and decay based on the new axiom
(\ref{9+}), (\ref{9-}). We exclude from our discussion here higher
order poles of the $S$-matrix (which are discussed in \cite{6}) and
other singularities.

The axiom (\ref{9+}), (\ref{9-}) has been postulated as a replacement
of the Hilbert space axiom in order to accommodate these new vectors
$|E^-\>\in\Phi_+^\times$, $|E^+\>\in\Phi_-^\times$ and the Gamow
vectors \cite{8}, $|z_R^-\>\in\Phi_+^\times$.  This replacement of
\eqref{8} by \eqref{9+}, \eqref{9-} is the only modification of the
traditional formulation~\cite{4}, that uses Hilbert space and Dirac
kets of~\eqref{8}. All other axioms of quantum theory remain intact, but these
other axioms are extended to the new objects like the Gamow kets
$|z^-_R\>$ and the kets $|E-i\epsilon^-\>$, the bras
$\<^+E-i\epsilon|$ and their analytic continuations $|z^-\>$,
$\<^+z|$, which we shall call Lippmann-Schwinger kets \cite{10b}
because of their analogy.  In particular, the fundamental Born
probabilities of Axiom \eqref{5} for an observable
$|\psi_\eta^-(t)\>\<\psi_\eta^-(t)|$ in the state $\phi^+$,
$|\<\psi_\eta^-(t)|\phi^+\>|^2$ are extended to Gamow states:
$\phi^+\rightarrow|z_R^-\>$. Thus the probability to detect the
observable $|\psi_-^\eta(t)\>\<\psi_-^\eta(t)|$ in the Gamow state
$|z_R^-\>$ is in analogy to \eqref{5}, with $\phi^D\rightarrow
|z_R^-\>$, given by:
\begin{equation}
\label{n14}
{\cal P}_\eta(t) \sim \text{Tr}\left(
  |\psi_\eta^-(t)\>\<\psi_\eta^-(t)|\,|z_R^-\>\<^-z_R|\right)
= |\<\psi_\eta^-(t)|z_R^-\>|^2 \sim |\<\psi_\eta^-(t)|\phi^G\>|^2\,.
\end{equation}

These new probabilities like~\eqref{n14} are mathematically well
defined quantities, (i.e., the value of the functional $|z^-_R\>\in
\Phi^\times_+$ at the point $\psi^-\in \Phi_+$) and represent Born
probabilities for the decay products $\eta$ at time $t$ in the ``Gamow
state'' $\phi^G\sim|z_R^-\>$.

The Gamow vectors $|z_R^-\>$ are the generalized eigenvectors or
eigenkets of the self-adjoint Hamiltonian $H$ with complex eigenvalue
$z_R$, they are associated to the first order pole of the $S$-matrix.
They have a Breit-Wigner energy distribution \eqref{2} and an
exponential time evolution. We give a brief sketch to show how this
follows from the new axiom (\ref{9+}, \ref{9-}), for details see
\cite{7,X}.

Since the Hardy space triples (\ref{9-}) and (\ref{9+}) are Rigged
Hilbert spaces, one has for every $\phi^\pm\in\Phi_\mp$ the Dirac
basis vector expansion (nuclear spectral theorem)
\begin{equation}
\phi^\pm=\sum_{j,\eta}\int^{\infty}_{0}dE|E,j,\cdots,\eta^\pm\>\<^\pm
E,j,\cdots,\eta|\phi^\pm\>\,.\label{13}
\end{equation}

Under the Hardy space hypothesis \eqref{9+}, \eqref{9-} the contour of
integration in (\ref{13}) can be deformed into the complex semi-plane
$C_-$ (in \eqref{13}, $\cdots$ are additional quantum numbers like
angular momentum $j\,,j_3$ etc, and $\eta$ denotes the channel or
particle species label.)

The starting point for the derivation of the Gamow ket is the Born
probability amplitude for the observable $\psi^-\in \Phi_+$ in the
state $\phi^+\in \Phi_-$, i.e. the $S$-matrix element
\begin{equation}
(\psi^{\textrm{out}},S\phi^{\textrm{in}})=(\psi^-_\eta,\phi^+_{\eta^0})=
\sum_j
\int^{\infty}_{0}
dE\<\psi^-_\eta|E^-\>S_j^{\eta\,\eta_0}(E)\<^- E|\phi^+_{\eta^0}\>.\label{14}
\end{equation}

The right hand side of \eqref{14} is obtained from \eqref{13} with
\begin{equation}
S^{\eta,\eta^0}_{j}(E)\equiv \<^- E,j,\eta|E,j\eta^0\cdots ^+\>
=2ia^\eta_j(E)\quad(\eta\ne\eta_0)\,.\label{15'}
\end{equation}
and using (rotation) invariance, for details see \cite{7,X}.

Under the Hardy class hypothesis (\ref{9+}, \ref{9-}) for the
wavefunctions $\<E^-|\psi^-\>$ and $\<E^+|\psi^+\>$, the integral in
(\ref{14}) can be carried out over any contour in the lower half
complex plane 2nd sheet of the $S$-matrix as long as it avoids
singularities.  If this integration is performed on a contour that
extends beyond the position $z_R$ of the resonance pole, then one has
to make a circle around it.

To the integral around this circle only the pole term
$\frac{R_{-1}}{z-z_R}$ of the $S$-matrix
\begin{equation}
S^{\eta,\eta'}_{j}(z)=\frac{R_{-1}}{z-z_R}+R_0+R_1(z-z_R)+\cdots\label{15}
\end{equation}
contributes and this contribution is according to the Cauchy integral
formula given by
\[
-2\pi i R_{-1}\<\psi^-|z^-_R\>\<^ + z_R|\phi^+\>\,.
\]
Since $\psi^-$ in \eqref{14} is arbitrary, we conclude that $|z^-_R\>$
is a functional on $\Phi_+=\{\psi^-\}$. The value of this functional
at $\psi^-\in\Phi_+$ is given by
\begin{equation}
\<\psi^-|z^-_R\>=\frac{i}{2\pi}\oint
dz\frac{\<\psi^-|z^-\>}{z-z_R}=\frac{i}{2\pi}\int^{+\infty}_{-\infty_{II}}
dE\frac{\<\psi^-|E^-\>}{E-z_R}\,,\label{16a}
\end{equation}
where $z_R=E_R-i\Gamma_R/2$ is the position of the resonance pole of
the analytic S-matrix. The first equality in (\ref{16a}) is the Cauchy
integral formula for the function $\<\psi^-|z^-\>$ and the second
equality is the Titchmarsh theorem for Hardy functions
$\<\psi^-|E^-\>$. The integral extends along the real axis in the 2nd
sheet as indicated by $-\infty_{II}$.

This equation (\ref{16a}) one also writes in Dirac notation as an
equation between functionals (omitting the arbitrary $\psi^-\in
\Phi_+$); one defines a ``normalized'' Gamow ket
$\phi^G_j\in\Phi_+^\times$:
\begin{equation}
\phi^G_j=\frac{\sqrt{2\pi \Gamma}}{f}|z_R,j,\cdots ^-\>=\frac{i\sqrt{2\pi
\Gamma}}{2\pi f}\int^{+\infty}_{-\infty_{II}
}dE\frac{|E_j,\cdots^-\>}{E-z_R}\,,\label{16b}
\end{equation}
where $f$ is a suitable "normalization" factor for the Gamow vector
$\phi^G_j$. Gamow kets $|z_R,j,\cdots^-\>$ are thus the singular
points of the analytically continued Lippmann-Schwinger kets
$|E,j,\cdots^-\>$ associated to the state given by the pole at $z_R$.

For the vector $\phi^G_j\in \Phi^\times$ defined in (\ref{16b}) and
{\em only} if the integral in \eqref{16b} extends to $-\infty_{II}$
one can derive (using the Hardy hypothesis (\ref{9-})) that
\begin{equation}
\<H\psi^-_\eta|\phi^G\>\equiv
\<\psi^-_\eta|H^\times|\phi^G\>
=(E_R-i\Gamma/2)\<\psi^-_\eta|\phi^G\>\,\,\textrm{for
all}\,\, \psi^-_\eta\in \Phi_+\label{17}
\end{equation}
where $H=H_0+V$ is self-adjoint (and semi-bounded~\footnote{The
  spectrum of a Hamiltonian is always bounded from below(``stability
  of matter''); in \eqref{14} we chose this bound to be zero and
  ignored bound state poles which are not relevant for our discussion
  here}). The result \eqref{17} justifies the notation
\[
\phi^G_j=\sqrt{2\pi \Gamma}|E_R-i\Gamma/2,\cdots ^-\>.
\]

In Dirac's notation the arbitrary $\psi^-_j\in \Phi_+$ is omitted and
\eqref{17} is written as
\begin{equation}
H^\times|E_R-i\Gamma/2, j,\cdots^- \>=(E_R-i\Gamma/2)\,\,|E_R-i\Gamma/2,
j,\cdots^-\>\,.\label{18}
\end{equation}
The operator $H^\times$ is uniquely defined by:
\begin{equation*}
\<H\psi^-|F^-\> = \<\psi^-|H^\times|F^-\>\quad\text{for every}\quad
F^-\in\Phi_+^\times\,,\quad \psi^-\in\Phi_+\,;
\end{equation*}
it is the extension of the operator $H^\dag ={H}$ to $\Phi^\times_+$.
For the Gel'fand triplet over Schwartz space \eqref{8}, self-adjoint
operators $H$ have only real generalized eigenvalues. For other
Gel'fand triplets, like the Hardy triplets \eqref{9+} and \eqref{9-},
the generalized eigenvalues of self-adjoint ${H}$ can also be complex.
(Dirac also omitted the $^\times$ of $H^\times$ when he wrote the
eigenvalue equation \eqref{18} for real generalized eigenvalues.)

An important result that one can derive for the vector $\phi^G_j$
defined by \eqref{16a} or \eqref{16b} is \cite{7,X}
\begin{multline}
\<\psi^-_\eta(t)|\phi^G_j\>\sim
\<e^{iHt/\hbar}\psi_\eta^-|E_R-i\Gamma/2^-\>\\ \equiv
\<\psi^-_\eta|e^{-iH^\times t/\hbar}|E_R-i\Gamma/2^-\>
=e^{-iEt/\hbar}e^{-\frac{\Gamma}{2}t/\hbar}
\<\psi^-_\eta|E_R-i\Gamma/2^-\>\qquad\\ \forall
\quad \psi^-_\eta\in \Phi_+\quad \text{and \fbox{for $\displaystyle
    t\ge 0$ only}}\,.\label{19}
\end{multline}

\section{Quantum mechanical time asymmetry}
The importance of the result \eqref{19} is that it can be obtained
from \eqref{16a} with $\psi^-\rightarrow \psi_\eta^-(t)$ only for
$t\ge 0$, since $\psi_\eta^-(t) = e^{iHt/\hbar}\psi_\eta^-$ is not an
element of $\Phi_+$ for $t < 0$.  The result \eqref{19} written as a
functional equation for the Gamow ket is:
\begin{equation}
\phi^G_j(t)=e^{-iH^\times t/\hbar}\phi^G_j=e^{-iEt/\hbar}e^{-(\Gamma
  /2) t/\hbar}\phi^G_j\quad
\text{for }t\ge 0\quad\text{only}\,.\label{20}
\end{equation}
This means that the Gamow ket \eqref{16b} defined as a functional
\eqref{16a}, \eqref{16b}, with Breit-Wigner energy distribution
$\frac{1}{E-z_R} = \frac{1}{E-(E_R-i\Gamma/2)}$ has a time evolution
\eqref{19} for $t\ge 0$ only. The Gamow kets~\eqref{20} are defined as
functionals for $t\geq0$ only.  The time evolution operators form a
semigroup $U^\times(t) = e^{-iH^\times t/\hbar}$, $t\ge 0$, and
\textit{not} a unitary group like the time evolution in the Hilbert
space given by: $U^\dagger(t) = e^{-i{H}t/\hbar}$, $-\infty < t <
\infty$, with the inverse $(U^\dagger(t))^{-1} = U^\dagger(-t)$. The
operator $U^\times(t) = e^{-iH^\times t/\hbar}$ with $0\le t < \infty$
has no inverse, hence $\{U^\times(t)| 0 \le t < \infty\}$ form a
semigroup.

Since $\<\psi^-_\eta|\phi^G(t)\>$ represents according to \eqref{n14}
the probability amplitude for the decay product $\eta$ (described by
$\psi^-_\eta$) in the state $\phi^G(t)$ we have derived the
exponential law:
\begin{equation}
|\<\psi^-_\eta(t)|\phi^G\>|^2=|\<\psi^-_\eta|\phi^G(t)\>|^2=e^{-\Gamma t/\hbar}
|\<\psi^-_\eta|\phi^G_j(0)\>|^2,\quad\textrm{for}\quad t\ge 0\,.\label{21}
\end{equation}

This is the exponential law which leads to the exponential decay rate
formula \eqref{4} of $R_\eta(t)$, by which the lifetime is measured
iff $\tau = \hbar/\Gamma$, where $\Gamma = -2\text{Im}(z_R)$ is the
width of the Breit-Wigner resonance \eqref{2}.

To summarize, the pole of the $j$-th partial $S$-matrix $S_j(z)$ in
the lower half complex energy semi-plane second sheet at $z_R =
E_R-i\Gamma/2$, the first term of \eqref{15}, defines the resonance
with resonance energy $E_R$ and resonance width $\Gamma$. From this
resonance pole one obtains a Gamow vector by \eqref{16a} and
\eqref{16b}. This Gamow vector is an eigenket of the Hamiltonian
fulfilling \eqref{17}, \eqref{18}. It describes an exponentially
decaying state with a lifetime precisely given by $\tau =
\frac{\hbar}{\Gamma}$.

This Gamow vector, defined by~\eqref{16a}, \eqref{16b} from the
S-matrix pole, has an energy distribution given by the ``idealized''
Breit-Wigner (Lorentzian)
\begin{equation}
\label{lor}
\frac{1}{E-(E_R-i\Gamma/2)}\qquad -\infty < E < \infty\,,
\end{equation}
that extends over the entire real axis, whereas the (Hilbert space)
spectrum of $\bar{H}$ is $0\le E<\infty$.

The Gamow vector $|z_{R}^{-}\rangle\in\Phi_{+}^{\times}$ is a
generalized vector (ket) on the Hardy space $\Phi_{+}$, which is
isomorphic (algebraically and topologically) to the space of wave
functions $\{\langle^{+}E|\phi^{+}\rangle\} ={\cal S}\cap{\cal
  H}_{-}^{2}|_{\mathbb{R}_{+}} =\{\langle\psi^{-}|E^{-}\rangle
=\overline{\langle^{-}E|\psi^{-}\rangle}\}$ analytic in the lower
complex semi-plane. The Hardy function $\langle^{+}z|\phi^{+}\rangle$
(in particular the $\langle^{+}E|\phi^{+}$ for $E\in\mathbb{R}_{-}$)
is already completely determined by its values for
$E\in\mathbb{R}_{+}$~\cite{winter}. The Hardy space vectors
$\phi^{+}\in\Phi_{-}$ (\textit{i.e.} $\langle^{+}E|\phi^{+}\rangle \in
{\cal S}\cap{\cal H}_{-}^{2}|_{\mathbb{R}_{+}}$) are represented by
  smooth well behaved functions $\langle^{+}E|\phi^{+}\rangle$ in the
  spectral resolution
\[
\phi^{+}=\int_{0}^{\infty}dE'\,
|E'^{+}\rangle\langle^{+}E'|\phi^{+}\rangle.
\] 
However the Gamow ket $\phi^{G}\sim|z_{R}^{-}\rangle$ cannot be
represented in this way, in particular
\[
\phi^{G}\text{ is not } = \int_{0}^{\infty} dE' |E'^{+}\rangle
\frac{1}{E'-(E_{R}-i\frac{\Gamma}{2})}.
\]
But $\langle^{+}E|\phi^{G}\rangle$ is an intricate singular
expression~\cite{23a}. Nevertheless, $\phi^{G}$ has the
representation~\eqref{16a}, \eqref{16b} by a smooth
function~\eqref{lor} on the whole real line $\mathbb{R}$, with
$-\infty<E<+\infty$. It is this complicated relationship between the
exponential time dependence~\eqref{20} for the Gamow vector and the
Breit-Wigner energy distribution~\eqref{lor} on $\mathbb{R}$, but not
on $\mathbb{R}_{+}$ which made the uncovering of $\tau=\hbar/\Gamma$
as an exact relation so difficult.

The Hardy space axiom \eqref{9+}, \eqref{9-} provides a unified theory
of resonances and decay. This was the purpose for which the Hardy
space hypothesis was introduced \cite{7}, it relates quantum decay
with resonance scattering. Further, the Hardy space admits Gamow state
vectors with an exponential time evolution of lifetime $\tau$ (which
cannot exist in a Hilbert space \cite{9c}) and relates them to vectors
with an idealized Breit-Wigner energy distribution \eqref{lor} of
width $\Gamma$.  And it led to the lifetime-width relation $\tau =
\hbar/\Gamma$, which physicists always desired as an exact equality.
This equality has recently been verified with an accuracy that exceeds
the accuracy expected by the Weisskopf-Wigner approximation \cite{9a}.

For the unification of resonance and decay phenomena, the Hardy axiom
should be welcome.  However, there is another conclusion drawn from
hypothesis \eqref{9+} \eqref{9-} which is expressed in \eqref{19} and
\eqref{20} by the time asymmetry $t\ge 0$, and this is not an easily
acceptable feature. It is a mathematical consequence of the boundary
conditions $\psi^-\in \Phi_+$, $\phi^+ \in \Phi_-\subset
\Phi^\times_+$, $\phi^G\in\Phi^\times_+$ for the time symmetric
dynamical equation (the Heisenberg equation for $\psi^-$ and the
Schrodinger equation for $\phi^+$ and $\phi^G$) inherent in the
hypothesis \eqref{9+} \eqref{9-}.  Whereas the solutions of the
dynamical equations under the Hilbert space boundary conditions and
under \eqref{8} are given (according to the Stone-von Neumann theorem
\cite{8a}) by the unitary group
\begin{equation}
\phi(t)=e^{-i\bar{H}t/\hbar}\phi,\qquad -\infty<t<+\infty\qquad \textrm{(for
the Schrodinger equation)}\label{21'}
\end{equation}
and
\begin{equation}
\psi(t)=e^{i\bar{H}t/\hbar}\psi,\qquad -\infty<t<+\infty\qquad \textrm{(for
Heisenberg equation),}\label{22}
\end{equation}
the solutions of the same dynamical equation under the Hardy space
boundary conditions \eqref{9+} \eqref{9-} are given (according to the
Paley-Wiener theorem \cite{koosis}) by the semigroups:
\begin{equation}
\phi^+(t)=e^{-iH^\times t/\hbar}\phi^+,\qquad 0\le t<+\infty\qquad
\textrm{(for the Schrodinger equation)}\label{23}
\end{equation}
and
\begin{equation}
\psi^-(t)=e^{iHt/\hbar}\psi^-,\qquad 0\le t<+\infty\qquad \textrm{(for
  the Heisenberg equation).}\label{24}
\end{equation}

The results \eqref{23} \eqref{24} mean that the time evolution
operator $e^{iHt/\hbar}$ is a continuous operator (with respect to the
topology in $\Phi_\mp$) only for $t\ge 0$ and therefore the time
evolution operator $e^{-iH^\times t/\hbar}$ in $\Phi^\times_\mp$ is
defined only for $t\ge 0$.  In physical terms this means that only for
$t\ge 0$ will the Born probabilities \eqref{n14} always be finite. The
restriction of the time evolution to the semigroup $U(t) =
e^{iHt/\hbar}$ $(t\ge 0)$ for observables and to $U^\times(t) =
e^{-iH^\times t/\hbar}$ $(t\ge 0)$ for states, is the way how the
Hardy axiom avoids infinite Born probabilities (the ``exponential
catastrophe'') even though it admits such vectors like Gamow kets
$\phi^G$ to represent physical entities, like exponentially decaying
states.

For the interpretation in terms of states $\phi^+$ and observables
$\psi^-_\eta(t)$, time asymmetry $t\ge 0$ means that the state
$\phi^+$ must be prepared first, at a time $t_0=0$, before the decay
products represented by $|\psi^-_\eta(t)><\psi^-_\eta(t)|$ can be
registered by the $\eta$-detector at time $(t-t_0)>0$. This is a
reasonable {\em condition of causality}.

All these features are not contained in the traditional (Hilbert
space) quantum mechanics where the Born probabilities
$|(\psi(t),\phi)|^2$ are defined for all times $-\infty<t<+\infty$.
But the Hilbert space probabilities are not without pathologies: One
can show that they must be different from zero for all time, unless
they are identically zero \cite{9b}, i.e., there is no decay of a
state which had been prepared at a finite time $t_0$ and before which
time the probabilities were zero. Further there exists no state vector
$\phi$ with exponential Born probabilities \cite{9c}.

\section{Conclusion}
The irreversible nature of quantum decay which can be understood as a
consequence of the time asymmetry \eqref{23} \eqref{24}, has been
mentioned in textbooks \cite{6a,6b} and lecture notes \cite{10,11}.

The time $t_0$ before which ``the state is defined completely by the
preparation'' has already been mentioned by Feynmann \cite{feynmann};
and Gell-Mann and Hartle \cite{11} applied this idea to the
probabilities of history (for the expanding universe considered as a
closed quantum system). They did not derive \eqref{24} but restricted
by fiat the time in $e^{iH(t-t_0)}$ to $t\ge t_0=\text{big-bang}$.
Our universe considered as a quantum physical system (one specimen of
an ensemble of many worlds) could be in states $\rho(t) =
e^{-iH(t-t_0)}\rho(t_0)e^{iH(t-t_0)}$ only for $t\ge t_0$ $\equiv
t_\text{big-bang}$. The big-bang time would thus provide an example
for the semigroup time $t_0$. Other systems where one could get an
indication of the existence of the semigroup time $t_0$ is the slow
(weak) decay of quasi-stable particles produced by a strong
interactions \cite{12}.

In general, it is difficult to recognize the quantum mechanical
beginning of time $t_0$ since in micro physics one studies a large
ensemble of (identical) quantum systems prepared at a collection of
numerous times. This makes it impossible to pin-point a particular
time $t_0$ at which the quantum state has been prepared.

For this reason the existence of a quantum mechanical beginning time
$t_0$ remained obscure for long. However, recently, experiments with
single ions \cite{14} changed this situation.  We shall discuss in the
subsequent publication II how these experiments demonstrate the
beginning of time for a quantum state.

\section*{Acknowledgment}
This work is part of a collaboration with the University of Texas at
Austin sponsored by the US National Science Foundation Award No.
OISE-0421936.


\begin{thebibliography}{99}
  
\bibitem{1} V.~Weisskopf and E.P.~Wigner, Z.~f.~ Physik {\bf 63}, 54 (1930).
  
\bibitem{1a} W.~Hollik, private communication.

\bibitem{2} M.L.~Goldberger, K.M.~Watson, \textit{Collision, Theory} (Wiley,
  New York, 1964) which is a presentation and application of
  Weisskopf-Wigner approximate methods.
  
\bibitem{4} R.F.~Streater, A.S.~Wightman, {\em PCT, Spin and
    Statistics and All That}. Princeton University Press, 1980,
  Chapter 2. R. Haag, {\em Local Quantum Physics}, Springer , Berlin
  1992, Chapter II.

\bibitem{8} G.~Gamow, Zeitschr Phys. 51, (1928).
  
\bibitem{10a} G.~Breit, E.~P.~Wigner, Phys.~Rev. {\bf 49}, 519 (1936);
  V.~A.~Fock, {\em Fundamentals of Quantum Mechanics}, Moscow (1931).
  
\bibitem{10b} B.A.~Lippmann and J.~Schwinger, Phys.~Rev.~{\bf 79},
  469 (1950); M.~Gell-Mann and H.~Goldberger, Phys.~Rev. {\bf 91}, 398
  (1953); W.~Brenig and R.~Haag, Fortschr.~Phys.~{\bf 7}, 183 (1959).
  
\bibitem{10c} C.~Moller, K.~Danske, Vid.~Selsk.~{\bf 22}, 19 (1946);
  R.J.~Eden, P.V.~Landshoff, P.J.~Olive and J.C.~Polkinghorne, {\em
    The Analytic S-Matrix} (Cambridge University Press, Cambridge,
  1966).

\bibitem{9c} L.A.~Khalfin, Sov.\ Phys.\ Dokl. {\bf 2}, 340-344 (1957);
  Sov Phys. JETB {\bf 6}, 1053 (1958).
  
\bibitem{10e} A.~Bohm, M.~Gadella, G.B.~Mainland, Am.~J.~Phys.~{\bf
    57}, 1103 (1989).

%\bibitem{9b} G.~C.~Hegerfeldt, Phys Rev. Lett. {\bf 72}, 596 (1994); {\em Irreversibility and
%Causality in Quantum Theory Semigroups and Rigged Hilbert Space}, Vol. 504,
%Springer Lecture Notes in Physics, edited dy A. Bohm, H.- D. Doebner, P.
%Kielanowski, (Springer, Berlin, 1998 ), p.~238.
  
\bibitem{9b} A.~Bohm, N.L.~Harshman, H.~Walther, Phys.\ Rev.~A {\bf
    66}, 012107 (2002).
  
\bibitem{3} M.~Levy, ``there does not exist a rigorous theory of which
  these various (Weisskopf-Wigner) methods can be considered as
  approximations'', \textit{On the Description of Unstable Particles in Quantum
  Field Theory}, Nuovo Cimento. {\bf 13}, 115-143 (1959).
  
\bibitem{6a} E.~Merzbacher, \textit{Quantum Mechanics}, Wiley N.Y.
  (1970), Chapter~18; C.~Cohen-Tannoudji, B.~Diu and F.~Laloe, {\em
    Quantum Mechanics}, Volume II (Wiley, New York, 1977), p.~1345,
  1353-54.
  
\bibitem{6b} T.D.~Lee, {\em Particle Physics and Introduction to
    Field Theory}, Chapter 13, (Harwood Academic, New York, 1981).
  
\bibitem{8a} M.H.~Stone and J.~von Neumann, Ann.\ Math. {\bf 33}, 567
  (1932).
  
\bibitem{8b} I.M.~Gel'fand and N.J.~Vilenkin, {\em Generalized
    Functions IV}, Academic Press, N.~Y. (1967).
  
\bibitem{8c} K.~Maurin, {\em Generalized Eigenfunction Expansions and
    Unitary Representations of Topological Groups}, Polish Scientific
  Publishers PWN, Warsaw (1968).
  
\bibitem{7} A.~Bohm, S.~Maxson, M.~Loewe, M.~Gadella; Physica A {\bf
    236}, 485 (1997); A.~Bohm, J.\ Math.\ Phys. {\bf 22}, 2813 (1981);
  Lett.\ Math.\ Phys. {\bf 3}, 455 (1978).
  
\bibitem{5} A.~Bohm, {\em Quantum Mechanics Foundations and
    Applications}, Section XXI.4, Springer N.Y.  (1979), 3rd edition,
  3rd printing, Springer , New York (2001).  A.~Bohm and M.~Gadella,
  \textit{Dirac Kets, Gamow Vectors and Gel' fand Triplets}, Lecture
  Notes in Physics, Volume 348, (Springer-Verlag, Berlin, 1989). See
  also Appendix of \cite{7}.

  
\bibitem{9} M.~Gadella, J.\ Math.\ Phys. {\bf 24}, 1462 (1983).
  
\bibitem{6} L.~Antoniou, M.~Gadella, G.~Pronko, J.\ Math.\ Phys. {\bf
    39}, 2459 (1998); A.~Bohm, M.~Loewe, S.~Maxson, P.~Patuleanu,
  C.~Puntmann, M.~Gadella, J.\ Math.\ Phys. {\bf 38}, 1 (1997).
  
\bibitem{X} Arno R.~Bohm, Proceedings of the 5-th Workshop of Time
  Asymmetric Quantum Theory, Trieste (2002), Intern.\ J.\ Theor.\ 
  Phys.\ \textbf{42}, 2317--2338 (2003).
  
\bibitem{9a} The relation $\tau = \Gamma/\hbar$ has been tested to an
  accuracy that goes beyond the Weisskopf-Wigner approximation only
  for the 3 $^2P_{3/2}$ of Na, C.W.~Oates, K.R.~Vogel, and
  J.L.~Hall, \textit{High Precision Linewidth Measurement of Laser-Cooled
  Atoms: Resolution of the Na $3P^2 P_{3/2}$ Lifetime Discrepancy},
  Phys.\ Rev.\ Lett. {\bf 76}, 2886-2865 (1996); U.~Volz, M.~Majerus,
  H.~Liebel, A.~Schmitt, and H.~Schmoranzer, {\em Precision Lifetime
    Measurements on NaI $3p^2P_{3/2}$ by Beam-Gas-Laser Spectroscopy},
  Phys.\ Rev.\ Lett.  {\bf 76}, 2862 (1996).
  
\bibitem{winter} C.~van~Winter, J.\ Math.\ Anal.\ Appl. {\bf 47}, 633
  (1974).
  
\bibitem{23a} I.~Antoniou, Z.~Suchanecki, S.~Tasaki in
  \textit{Generalized Functions, Operator Theory and Dynamical
    Systems}, Chapman and Hall Research Notes in Mathematics,
  I.~Antoniou and G.~Lumer (editors), vol.~399, page 130--143 (1999).
  
\bibitem{koosis} P.~Koosis, {\em Introduction to $H_p$ spaces}, page
  $130$, Cambridge University Press (1998).
  
\bibitem{10}
%E. g, E. Merzbacher, Quantum Mechanics, Wily N. Y. (1970) chap. 18;C.
%Cohen- Tannoudji, B. Din and F. Laloe, Quantum Mechanics, Volume ll (Wiley,
%New York, 1997), P. 1345, 1353-54; T. D. Lee, Particle Physics and
%Introduction to Field
%          Theory, Chapter 13, (Herwood Academic, New York, 1981); 
  I.~Antoniou, Proc. 2nd Internat. Wigner Symposium, Goslar 1991, Eds.
  H.-D. Doebner \textit{et. al}.  (World Scientific, Singapore, 1992);
  I.~Antoniou and I.~Prigogine, Physica A. {\bf 192}, 443 (1993);
  R.~Haag, Comm. Math. Phys. {\bf 132}, 245 (1990); R.~Haag, Lectures
  at the Max Born Symposium on" Quantum Future", Przieka (1997); N.
  van Kampen, Proceedings of the XXI Solvay Conference 1999,
  Ad.\ Chem.\ Phys. {\bf 122}, 301 (2002).
  
\bibitem{11} M.~Gell-Mann and J.B.~Hartle, in \textit{Physical Origins
    of Time Asymmetry}, J.J.~Halliwell et. al. (eds.), Cambridge
  University, Cambridge, 1994; M.~Gell-Mann and J.B.~Hartle,
  University of California at Santa Berbara, Report No. UCSBTH-95-28
  (1995), LANL Archives, gr-qc/9509054 and references thereof.
  
\bibitem{feynmann} R.~Feynmann, Rev.\ Mod.\ Phys. {\bf 20}, 369 (1948),
  see e.g., p.~372 and 379.
  
\bibitem{12} A.~Bohm, Phys.\ Rev.~A {\bf 60}, 861 (1999).
  
\bibitem{14} W.~Nagourney, J.~Sandberg, and H.~Dehmelt, Phys.\ Rev.\ 
  Lett. {\bf 56}, 2797 (1986); X.~Zhao, N.~Yu, H.~Dehmelt and
  W.~Nagourney, Phys.\ Rev.~A {\bf 51}, 4483 (1995); Th.~Sauter,
  W.~Neuhauser, R.~Blatt, and P.E.~Toschek, Phys.\ Rev.\ Lett. {\bf
    57}, 1696 (1986); J.C.~Bergquist, R.G.~Hulet, W.M.~Itano, and
  D.J.~Wineland, Phys.\ Rev.\ Lett. {\bf 57}, 1669 (1986); E.~Peik,
  G.~Hollemann, and H.~Walther, Phys.\ Rev.~A {\bf 49}, 402 (1994).

\end{thebibliography}
 \end{document}